\documentclass[aps,prx,twocolumn,longbibliography,amsmath,amssymb,nofootinbib,longbibliography,10pt]{revtex4-1}
\usepackage{amsfonts,amssymb,amsmath}
\usepackage{graphics}
\usepackage{graphicx}
\usepackage{mathrsfs}
\usepackage{mathtools}
\usepackage{textcomp}
\usepackage{verbatim}
\usepackage{bm}          
\usepackage{soul}
\usepackage{cancel}
\usepackage{upgreek}
\usepackage{txfonts}
\usepackage[colorlinks, linkcolor=blue]{hyperref}
\usepackage{color} 
\usepackage[papersize={8.5in,11in}]{geometry}
\usepackage{esint}
\newcommand{\be}{\begin{equation}}
\newcommand{\ee}{\end{equation}}
\newcommand{\e}{\varepsilon}

\newcommand{\q}{{\bf q}}
\newcommand{\kb}{k_\text{B}}
\newcommand{\R}{{\bf R}}

\newcommand{\abo}{a_B^{(0)}}

\newcommand{\ang}{\mathring{\textrm{A}}}
\renewcommand{\vec}[1]{\boldsymbol{#1}}
\def \k {{\vec k}}

\def \e {\epsilon}

\def \r {{\vec r}}
\def \R {{\vec R}}

\def \q {{\vec q}}

\def \beq {\begin{eqnarray}}
\def \eeq {\end{eqnarray}}

\usepackage{makecell}

\begin{document}

\title{Properties of the donor impurity band in mixed valence insulators}

\author{Brian Skinner}
\affiliation{Department of Physics, Massachusetts Institute of Technology, Cambridge MA 02139, U.S.A.}
\affiliation{Department of Physics, Ohio State University, Columbus, OH 43210, USA}
\date{\today}
\begin{abstract}

In traditional semiconductors with large effective Bohr radius, an electron donor creates a hydrogen-like bound state just below the conduction band edge.  The properties of the impurity band arising from such hydrogenic impurities have been studied extensively during the last 70 years.
In this paper we consider whether a similar bound state and a similar impurity band can exist in mixed-valence insulators, where the gap arises at low temperature due to strong electron-electron interactions.
We find that the structure of the hybridized conduction band leads to an unusual bound state that can be described using the physics of the one-dimensional hydrogen atom. The properties of the resulting impurity band are also modified in a number of ways relative to the traditional semiconductor case; most notably, the impurity band can hold a much larger concentration without inducing an insulator-to-metal transition.  We estimate the critical doping associated with this transition, and then proceed to calculate the dc and ac conductivities and the specific heat.  We discuss our results in light of recent measurements on the mixed-valence insulator SmB$_6$, and find them to be consistent with the experiments.

\end{abstract}
\maketitle

\section{Introduction}

In a mixed valence metal, a heavy $f$-like band of electronic states coexists with a lighter $s$- or $d$-like band at the Fermi level. \cite{MottMV,CMV76}  In some mixed-valence compounds, however, electron-electron interactions lead to the opening of a charge gap at low temperature, resulting in an insulating state which can be called a mixed-valence insulator (MVI).  The nature of this insulating state has attracted much attention in recent years, following a proposal that some MVIs could host an interaction-induced three-dimensional topological insulator state \cite{Coleman1,Kane10,Zhang11}, and the subsequent experimental observation of quantum oscillations coexisting with the insulating state in the MVIs SmB$_6$ \cite{SS15, Li16} and YbB$_{12}$ \cite{SSybb12, lim}.  

In both of these latter materials, measurements of the bulk resistivity at low temperature suggest an activation energy on the order of several meV, \cite{Batlogg1,paglione1,fisk1,fisk2,mcqueen1,balakrishnan,TKIrev, eo2018robustness} despite an (indirect) hybridization band gap that is on the order of 20 meV. \cite{Rossler_hybridization_2014, ruan_emergence_2014, paglione1, xu_surface_2013, Neupane_surface_2013, frantzeskakis_kondo_2013} In a conventional semiconductor, such a reduced activation energy could be understood as a consequence of mid-gap donor impurity states, which appear below the conduction band edge and tend to pin the chemical potential to partially-filled impurity bands.  The usual description of these impurity states makes use of the effective mass approximation for describing states near the bottom of the conduction band, and gives a hydrogen-like quantum state with effective Bohr radius
\beq 
\abo = \frac{\hbar^2 \e}{m e^2} = (0.53\, \ang) \frac{\e}{m/m_0}
\label{eq:ab}
\eeq
(in Gaussian units)
and an ionization energy
\beq
E_i = \frac{e^2}{2 \e \abo} = \frac{m e^4}{2 \e^2 \hbar^2} = (13.6 \text{\,eV}) \frac{m/m_0}{\e^2}.
\label{eq:Eb}
\eeq
This description is valid so long as the ionization energy $E_i$ is much smaller than the band gap $E_g$ and the Bohr radius $\abo$ is much longer than the lattice constant $a_0$.
Here, $m$ is the effective mass, $m_0$ is the bare electron mass, $\epsilon$ is the dielectric constant, $\hbar$ is the reduced Planck constant, and $-e$ is the electron charge.

Implicit in Eqs.\ (\ref{eq:ab}) and (\ref{eq:Eb}) is the assumption that the band edge is well described by a parabolic dispersion relation, $E(k) \propto k^2$, where $E$ is the quasiparticle energy and $k$ is the wave vector relative to the position of the band minimum in reciprocal space.  However (as pointed out in Ref.\ \onlinecite{rakoski_understanding_2017}), attempting to directly apply this description to MVIs can give nonsensical results, including an ionization energy that is either unreasonably large or unreasonably small, depending on whether one uses the heavy (``$f$-band'') mass or the light (``$d$-band'') mass.  The issue is that the conduction and valence bands at low temperature arise from hybridization between coexisting light and heavy bands, and the resulting hybridized bands have a ``Mexican hat" shape (as pointed out, for example, in Ref.\ \onlinecite{CooperExc}, and as we discuss in the following section).

In this paper we consider the general question of how one should think about donor or acceptor impurity states and the resulting impurity band in the presence of such hybridized bands.  We address this question by first solving for the ground state wave function and its ionization energy using the continuum approximation (Sec.\ \ref{sec:boundstate}).  We then use this solution to derive a number of properties of the impurity band, including the critical doping required for the insulator-to-metal transition (Sec.\ \ref{sec:IMT}), the DC conductivity (Sec.\ \ref{sec:DC}), the AC conductivity (Sec.\ \ref{sec:AC}), and the specific heat (Sec.\ \ref{sec:CV}).  We focus everywhere on the bulk behavior, for which the topology of the band structure and the (potential) edge states play no role.

While our results apply generically to MVIs, and even more generally to any gapped system with a Mexican hat-shaped dispersion relation, for definiteness we focus our discussion around SmB$_6$.  In this context our main results are as follows:

\begin{itemize}

\item
 While a naive application of Eq.\ (\ref{eq:Eb}) does not yield a sensible result for the impurity ionization energy in SmB$_6$, properly accounting for the hybridized band structure in the solution to the hydrogen-like Schrodinger equation gives an ionization energy on the order of several meV.  This energy scale is in line with the bulk activation energy seen experimentally. \cite{Batlogg1,paglione1,fisk1,fisk2,mcqueen1,balakrishnan,TKIrev, eo2018robustness}
 
\item 
Given the existence of such impurity states, one can wonder why SmB$_6$ exhibits a bulk insulating state at all, since the concentration of dopant impurities (including substitutional atoms of C \cite{mcqueen1} or Gd \cite{fuhrman_screened_2017} and Sm vacancies \cite{konovalova_effect_1982, phelan_chemistry_2016, valentine_breakdown_2016}) in the studied samples is as high as a few percent \cite{phelan_chemistry_2016} and the typical spacing between impurities is much smaller than the spatial extent of the impurity wave function (discussed below).  One would generically expect that such a heavily-doped sample will find itself on the conducting side of the Mott criterion for the insulator-to-metal transition (IMT).  We show, however, that the Mott criterion is strongly modified in MVIs due to the Mexican hat structure of the band edge. In particular, the resulting donor impurity wave functions are such that, even though their spatial extent is relative large, they have poor overlap with each other in the quantum mechanical sense.  This reduced overlap allows the impurity band to remain insulating even at quite high impurity concentration.

\item 
The reduced quantum overlap also strongly modifies the result for the optical conductivity $\sigma(\omega)$ relative to the conventional semiconductor case.  While $\sigma(\omega)$ remains linear in the frequency $\omega$, as in conventional semiconductor impurity bands \cite{shklovskii_zero-phonon_1981}, the coefficient of proportionality is parametrically smaller at a given impurity concentration. Numerical estimates for our calculated coefficient is in line with recent experiments \cite{laurita_anomalous_2016}.

\item 
The specific heat $C_V$ of the impurity band is associated with quasiclassical rearrangement of electrons among localized states with strong Coulomb repulsion.  In this sense the specific heat is similar to the conventional case, except that the impurity band can hold a much higher-than-usual concentration of impurities, which enables a large value of $C_V$.  We find $C_V \propto T$, up to a logarithmic coefficient, with a magnitude that is consistent with experiments.

\end{itemize}

It is worth noting that previous authors have described impurity states in Kondo insulators using a model of a missing electron in an Anderson lattice (e.g., Refs.\ \onlinecite{Schlottmann1, Schlottmann2, Riseborough}).  In contrast, our focus here is on states created by charged donor impurities, for which the impurity state energy is dominated by the Coulombic attraction of the donor electron to the impurity charge.

\section{The hybridized band structure}
\label{sec:bands}

\begin{figure}[htb]
\centering
\includegraphics[width=1.0 \columnwidth]{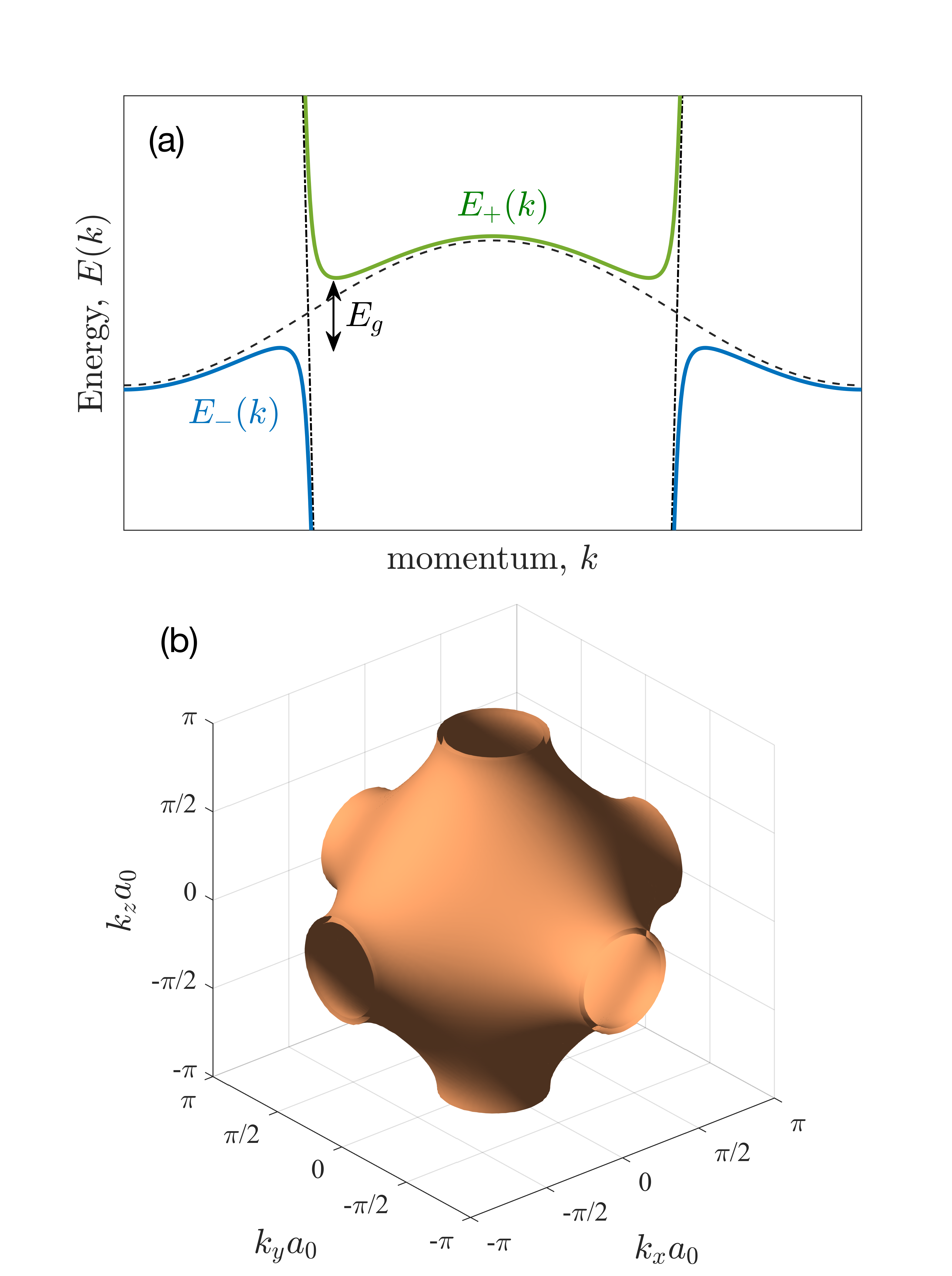}
\includegraphics[width=0.7 \columnwidth]{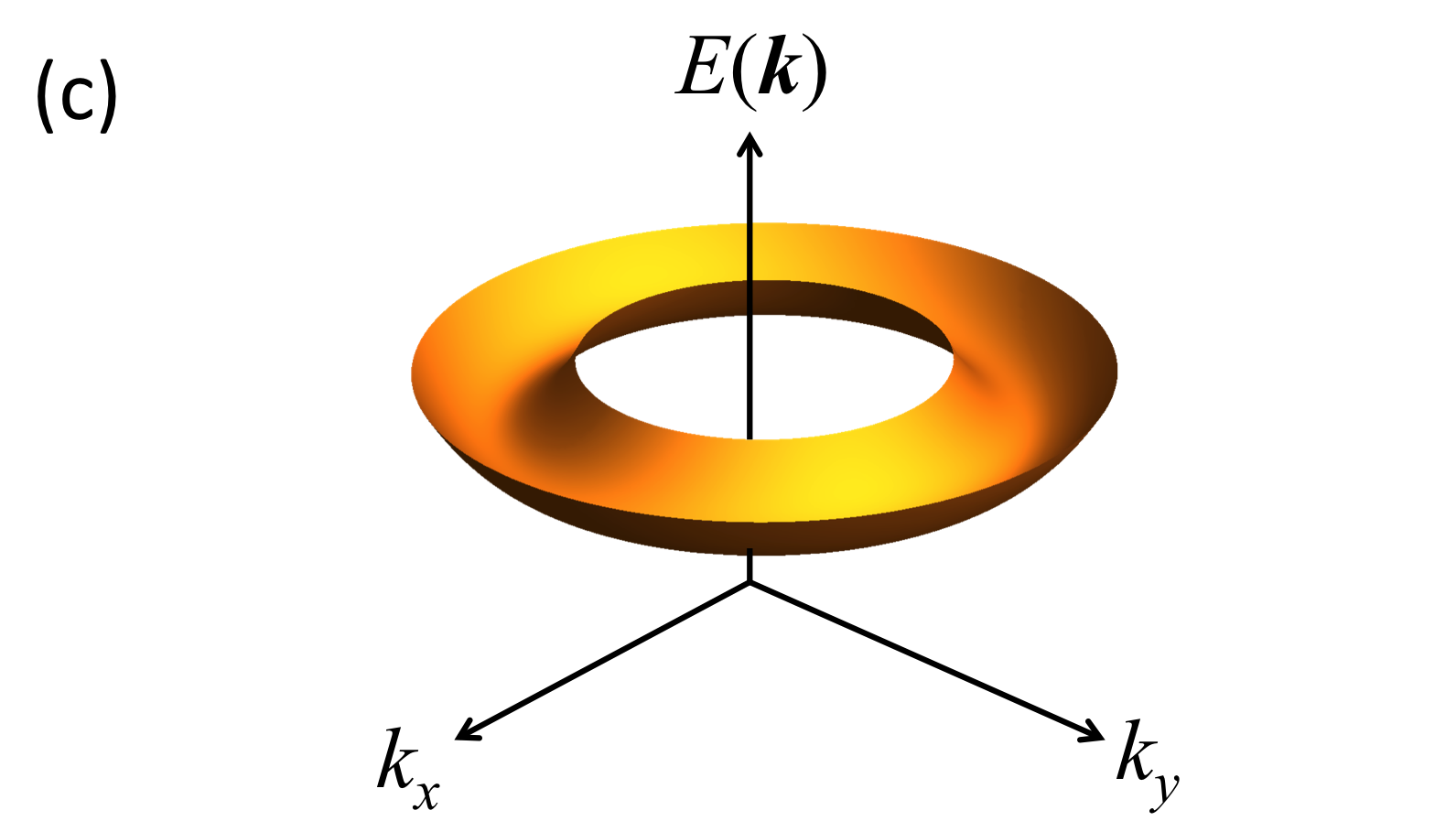}
\caption{(a) A schematic illustration of the dispersion relations $E_\pm(k)$ for the conduction and valence bands (solid lines) along some particular momentum direction in the vicinity of the band crossings.  The dashed and dash-dotted lines show the dispersion relations for the unhybridized $f$ and $d$ electrons, respectively. 
(b) The minimum energy of the conduction band, for which $E_{+}(\k) = E_g/2$, occupies a surface in the space of wave vector $\k$. 
(c) For most of this paper, we describe the conduction band edge using the simplified ``Mexican hat'' dispersion relation: $E_{+}(\k) = E_g/2 + \hbar^2(|\k| - k_0)^2/2m$.
}
\label{fig:dispersion}
\end{figure}

The generic description of hybridized bands $E_\pm (\k)$ arising from bare, unhybridized bands $E_d(\k)$ and $E_f(\k)$ is
\be
E_\pm(\k) = \frac{E_d(\k) + E_f(\k)}{2} \pm \sqrt{\left(\frac{E_d(\k)-E_f(\k)}{2}\right)^2  + |V(\k)|^2}.\nonumber\\
\label{eq:Epm}
\ee
Here the subscripts $d$ and $f$ indicate that $E_d(\k)$ and $E_f(\k)$ describe the dispersion relations for the unhybridized $d$ and $f$ bands, respectively, while $V(\k)$ is the  hybridization matrix element.  The upper band $E_+(\k)$ describes the conduction band and $E_-(\k)$ is the valence band [see Fig.\ \ref{fig:dispersion}(a)].  For concreteness, one can take the unhybridized dispersion relations $E_d(\k)$ and $E_f(\k)$ to be described by nearest-neighbor hopping on the cubic lattice (see Appendix \ref{sec:epsilon} for details, and the end of this section for further discussion regarding SmB$_6$).
The corresponding indirect band gap has width
\be 
E_g = 4 |V| \frac{\sqrt{m_d m_f}}{m_d + m_f},
\label{eq:Eg}
\ee
where $m_d$ and $m_f$ are the $d$- and $f$-band masses, respectively.  The gap associated with zero-momentum optical transitions is $2V$.  (Here we have replaced the function $|V(\k)|$ with the characteristic magnitude $V$ of the hybridization matrix element near the band crossing; see, for example, Ref.\ \onlinecite{fuhrman_ingap_2014} or the review in Ref.\ \onlinecite{TKIrev} for a more thorough discussion.)

For shallow impurity states, the effective mass is determined by the dispersion of the conduction band near the band minimum.  For the dispersion relations $E_d(\k)$ and $E_f(\k)$ that we have chosen, the conduction band minimum is not a point in momentum space, but a surface [as shown in Fig.\ \ref{fig:dispersion}(b)], so that the dispersion is flat in two orthogonal directions at any point in momentum space at the conduction band edge.  One can still define an effective mass $m$ by expanding to second order the conduction band energy in the direction perpendicular to the plane of the conduction band minimum. This expansion gives
\be 
\frac{m}{m_0} \simeq \frac{1}{8} \frac{V}{\hbar^2/(m_0 a_0^2)} \left(\frac{m_f}{m_0}\right)^{3/2} \sqrt{\frac{m_d}{m_0}}
\label{eq:meff}
\ee
in the limit $m_f \gg m_d$, where $a_0$ is the lattice constant.

In the remainder of this paper, we make an approximation in which the surface of conduction band minima is described as a sphere in momentum space rather than the more complicated shape shown in Fig.\ \ref{fig:dispersion}(b).  As we show below, the radius of the sphere enters the results for the ionization energy $E_i$ and the wave function size $a_B$ only in the argument of a logarithm, so that variations in the shape or size of the surface of minima do not appreciably alter our results. This spherical approximation is equivalent to taking the conduction band to have a ``Mexican hat'' dispersion relation
\be 
E_+(\k) \simeq \frac{E_g}{2} + \frac{\hbar^2}{2 m}(|\k| - k_0)^2
\label{eq:MHdispersion}
\ee
at low energies $0 < E_+ - E_g/2 \ll V$. This dispersion relation is illustrated in Fig.\ \ref{fig:dispersion}(c).
(The valence band has a similar, downward-facing dispersion.)  In MVIs such as SmB$_6$, the momentum scale $k_0$ is of order $\pi/(2 a_0)$.  

Future numerical estimates for SmB$_6$ require an estimate of the hybridization matrix element $V$.  Such an estimate can be made by examining the relation (\ref{eq:Eg}) between $V$ and the band gap $E_g$.  Experimental measurements of the low-temperature gap in SmB$_6$ using tunneling probes suggest that $E_g$ is on the order of $10$\,meV, \cite{Rossler_hybridization_2014, ruan_emergence_2014, paglione1} while optical probes observe a direct gap on the order of $20$\,meV \cite{xu_surface_2013, Neupane_surface_2013, frantzeskakis_kondo_2013}.  We choose $V = 15$\,meV, which gives a gap $E_g$ ranging between $5$\,meV and $20$\,mev, depending on the precise values chosen for the masses $m_d$ and $m_f$.\footnote{In SmB$_6$, photoemission \cite{DenlingerJPS} and optical conductivity \cite{TWoptics} studies give estimates for the $d$-band mass that range between $0.5m_0$ and $2.0m_0$. The heavy $f-$band mass, on the other hand, has a mass that is no smaller than $m_f \approx 15m_0$ and in some directions of momentum exceeds several hundred. \cite{DenlingerJPS, Denpvt, TWoptics}} 
Knowing the band structure also allows us to calculate self-consistently the dielectric constant $\e$; our choice $V = 15$\,meV gives $\e \approx 1600$, which is on the high side of the experimental estimates, while still providing a reasonable fit to both the optical and density of states gaps. Experimental estimates of the low-temperature dielectric constant in SmB$_6$ have given $\e \approx 600$ \cite{gorshunov_low-energy_1999, sluchanko_intragap_2000} and $\e \approx 1500$. \cite{TWoptics}   The calculation of the dielectric constant is discussed in more detail in Appendix \ref{sec:epsilon}.   

We note also that in SmB$_6$ there are next-neighbor and third-neighbor hopping terms that are important for the dispersion relation; for example, Ref.\ \onlinecite{fuhrman_ingap_2014} suggests that third-neighbor hopping plays a large role. Such non-nearest-neighbor hopping terms lead to a deformation in the shape of the surface of conduction band minima relative to what is plotted in Fig.\ \ref{fig:dispersion}(b), but they do not change the dimensionality of the surface or the scale of its radius. Thus, to within the accuracy of the approximations made in this paper, the ``Mexican hat'' description of Eq.\ (\ref{eq:MHdispersion}) remains valid.

\section{The impurity bound state}
\label{sec:boundstate}

One might naively think that the correct results for the Bohr radius and the ionization energy are obtained by inserting Eq.\ (\ref{eq:meff}) for the effective mass into the conventional formulas, Eq.\ (\ref{eq:ab}) and (\ref{eq:Eb}).  However, these formulas are appropriate only for a conduction band whose energy increases parabolically with momentum in all directions away from a single point. For MVIs, one can arrive at an answer by solving the Schr\"{o}dinger equation for an electron with kinetic energy described by Eq.\ (\ref{eq:MHdispersion}) in a Coulomb potential $-e^2/\epsilon r$.
This problem was solved for two-dimensional systems in Ref.\ \onlinecite{skinner_bound_2014} (following Ref.\ \onlinecite{Chaplik}), and the calculation can be generalized in a straightforward way as follows.  During the remainder of this paper we focus, for concreteness, on electron donors, but our analysis can be applied equally well to acceptor impurities also.

The Schr\"{o}dinger equation in position space is
\be 
\hat{E}_{+} \psi(\r) - \frac{e^2}{\epsilon r} \psi(\r) = \left(\frac{E_g}{2} - E_i \right) \psi(\r),
\label{eq:Schr}
\ee 
where $\psi(\r)$ is the wavefunction, with $r$ the distance from the donor impurity, and $\hat{E}_{+}$ is an operator corresponding to the dispersion of the conduction band.  Equation (\ref{eq:Schr}) can be written in momentum space
as
\be 
E_{+}(k) \widetilde{\psi}(k) - \int \frac{d^3 q}{(2 \pi)^3} \frac{4 \pi e^2}{\epsilon |\k - \q|^2} \widetilde{\psi}(q) = \left(\frac{E_g}{2} - E_i \right)  \widetilde{\psi}(k).
\label{eq:Schk}
\ee
Here we have made use of the fact that the ground state wave function is radially symmetric, so that $\widetilde{\psi}(\k) = \widetilde{\psi}(k)$.

In the limit of small ionization energy $E_i \ll \hbar^2 k_0^2/m$, the momentum space wave function $\widetilde{\psi}(k)$ is strongly peaked around $k = k_0$.  Thus, the integrand in Eq.\ (\ref{eq:Schk}) is appreciable only along a thin shell of radius $|\q| = k_0$ in momentum space.  For wave vectors $\k$ with $|k - k_0| \ll k_0$, this shell can be approximated as an infinite plane, and
\begin{align}
\int \frac{d^3 q}{(2 \pi)^3} \frac{4 \pi e^2}{\epsilon |\k - \q|^2} \widetilde{\psi}(q)
& \simeq \frac{e^2}{2 \pi^2 \epsilon} \int dq_r \int d^2 q_\perp \frac{ \widetilde{\psi}(q_r) }{(q_r-k)^2 + q_\perp^2} \nonumber \\
& \simeq \frac{e^2}{\pi \epsilon} \int d q_r  \widetilde{\psi} (q_r) \ln \left( \frac{k_0}{|k - q_r|} \right).
\end{align}
Here $q_r$ represents the integration variable for momentum in the radial direction and $q_\perp$ represents the momentum in the plane of the thin shell.  The integral over $q_\perp$ is truncated at $q_\perp = k_0$.

With these simplifications we can rewrite the Schr\"{o}dinger equation, Eq.\ (\ref{eq:Schk}), as 
\be
\frac{\hbar^2 \delta_k^2}{2 m^*} \widetilde{\psi}(\delta_k) - \frac{e^2}{\pi\epsilon } \int d\delta_q  \ln \left( \frac{k_0}{|\delta_k - \delta_q|} \right) \widetilde{\psi}(\delta_q) 
= E \widetilde{\psi}(\delta_k),
\label{eq:Sch-radial}
\ee
where here we have introduced the notation $\delta_k = k - k_0$ and $\delta_q = q - k_0$.  
Written in the form of Eq.\ (\ref{eq:Sch-radial}), the Schr\"{o}dinger equation is identical to that of the one-dimensional (1D) hydrogen atom \cite{Loudon1959odh}:
\be 
\frac{\hbar^2 k^2}{2 m_0}\widetilde{\psi}(k) - \frac{ e^2}{\pi \epsilon} \int dq \ln \left(\frac{1/\lambda}{|k - q|} \right) \widetilde{\psi}(q) = E \widetilde{\psi}(k),
\nonumber
\ee
where $m_0$ is the physical electron mass and $\lambda$ is some small-distance cutoff to the Coulomb potential.  (In the absence of such a cutoff, the ionization energy of the 1D hydrogen atom is logarithmically divergent \cite{Loudon1959odh}.)  

\begin{figure}[htb]
\centering
\includegraphics[width=1
\columnwidth]{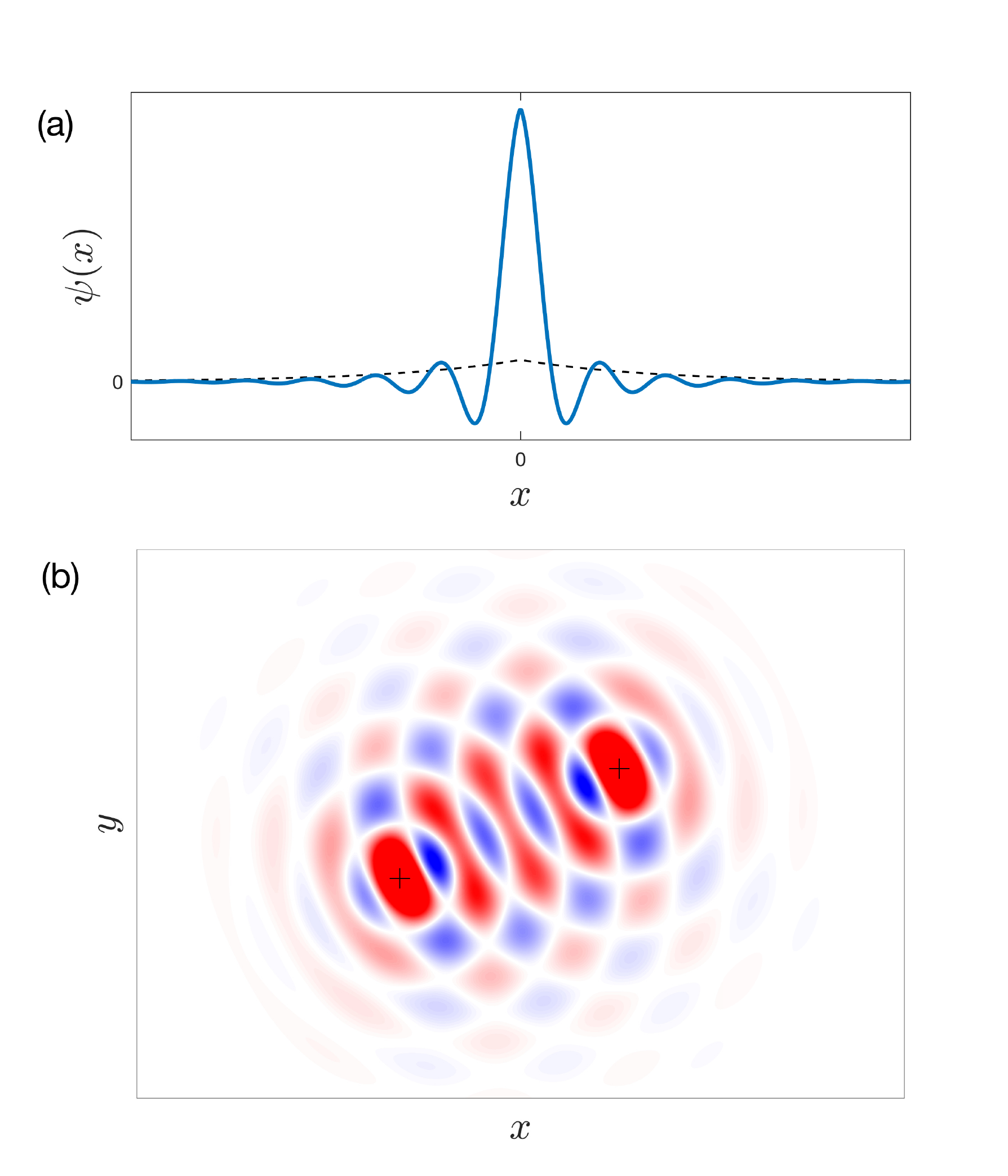}
\caption{(a) The ground state wave function of an electron bound to a donor impurity in a MVI, plotted as a function of some coordinate $x$ passing through the origin.  The dashed line shows the usual hydrogen-like result for conventional semiconductors, while the thick line shows the result of Eq.\ (\ref{eq:psi0}).
(b) Schematic depiction of the wave function overlap $\psi^*(|\r - \R|) \psi(r)$ between two impurity wave functions centered at different spatial locations (black $+$'s). The color denotes the value of the overlap, with red indicating positive and blue indicating negative. }
\label{fig:psi}
\end{figure}

The corresponding wavefunction is given by $\widetilde{\psi}(k) \propto [1 + a_B^2 \delta_k^2]^{-1}$, where 
\be 
a_B \approx \frac{\abo}{2 \ln[k_0 \abo]}.
\label{eq:aBMH}
\ee
and $\abo = \hbar^2 \epsilon/(m e^2)$ is the conventional Bohr radius.  This wave function $\widetilde{\psi}(k)$ corresponds to the 1D Fourier transform of the spatial wavefunction $\psi(x) \propto \exp[-|x|/a_B]$ for the 1D hydrogen atom, \cite{Loudon1959odh} with $k \rightarrow \delta_k$.  Taking the three-dimensional inverse Fourier transform of $\widetilde{\psi}(k)$ in the limit $k_0 a_B \gg 1$ gives for the bound state 
\be 
\psi(r) \simeq \frac{1}{\sqrt{\pi a_B}} \frac{\sin (k_0 r)}{r} \exp[-r/a_B].
\label{eq:psi0}
\ee
This wave function is plotted schematically in Fig.\ \ref{fig:psi}(a).
The corresponding ionization energy is
\be 
E_i \simeq \frac{2 e^2}{\epsilon \abo} \ln^2 [k_0 \abo].
\label{eq:EbMH}
\ee

Inserting parameters for SmB$_6$ into Eq.\ (\ref{eq:aBMH}) gives an effective Bohr radius $a_B$ that ranges between $40~\ang$ and $90~\ang$, while Eq.\ (\ref{eq:EbMH}) gives an ionization energy between $1$\,meV and $5$~meV, depending on the precise values of the band masses and the dielectric constant. Below we take $a_B = 60~\ang$ and $E_i = 2$\,meV for the purpose of making numerical estimates.  An alternative, variational estimate for the ionization energy that does not assume a spherically-symmetric dispersion relation is presented in Appendix \ref{sec:variation}.  This calculative gives a lower bound $E_i \approx 0.8$\,meV for the ionization energy in SmB$_6$.

It is worth noting that in our estimates of the ionization energy we have neglected the wave vector dependence of the dielectric function $\e(q)$.  This dependence provides a perturbation that strengthens the Coulomb potential at short distances, thereby further increasing the electron ionization energy.  The nature of this short-distance enhancement of the Coulomb attraction is generally dependent on lattice-scale details, and it is not well-described by the continuum approximation that we are employing here.

It is also important to note that we have restricted our attention to the case where the Kondo scale is much smaller than the band gap, so that the Kondo screening of mid-gap impurities by conduction band electrons can be neglected.  Other authors have examined the opposite limit, \cite{fuhrman2018magnetic} where Kondo screening is strong, motivated by the experimental observation that the magnetic moment of impurities can produce a large contribution to the specific heat. \cite{fuhrman_screened_2017}

\section{The insulator-to-metal transition}
\label{sec:IMT}

Let us now discuss the insulator-to-metal transition (IMT) in MVIs as a function of doping.  
In a traditional semiconductor with a parabolic dispersion relation, the critical concentration $N_c^{(0)}$ associated with the IMT is given by the Mott criterion, \cite{Mott1968}
\be 
N_c^{(0)} \approx (0.26/a_B)^3
\label{eq:Mottcrit}.
\ee 
Naively, one can think that this criterion implies that donor impurity states, which in conventional semiconductors have a typical spatial size $a_B = \abo$, must be strongly overlapping spatially in order to produce a conducting state.  For SmB$_6$, however, this Mott value $N_c$ is apparently as small as $10^{17}$\,cm$^{-3}$, or less than $0.001\%$, which is well below the level of uncontrolled doping in SmB$_6$. \cite{phelan_chemistry_2016}  Indeed, even SmB$_6$ samples with intentional doping as high as $5\%$ are seen to exhibit an insulating-like temperature dependence of the resistivity, \cite{fuhrman_screened_2017} which suggests that in SmB$_6$ the usual critical concentration $N_c$ arising from the Mott criterion should be replaced with a much larger one.

The classic paradigm for thinking about the IMT, as suggested by Mott and others (see, for example, Ref.\ \onlinecite{Mott1980}), can be summarized as follows.
On the insulating side of the transition, where donor impurities are sparse, the typical hopping integral $\gamma$ between neighboring impurity sites is small due to their large separation.  The on-site energy $U$ associated with double-occupation of a single donor is comparatively large.  The system therefore resembles a Mott insulator (on a spatially-irregular lattice). When the concentration of impurities is increased, the typical value of the hopping integral $\gamma$ is increased, and for concentrations greater than some critical concentration $N_c$ we have $\gamma > U$, which produces a Mott transition from an insulating to a metallic state. \cite{Mott1980}
In this section we extend this paradigm to the case of MVIs, and discuss its implication for the critical doping concentration $N_c$. 
We emphasize that our analysis does not constitute an authoritative theoretical treatment of the IMT in MVIs, but is best read only as a naive extension of established ideas for doped semiconductors.  
Our analysis suggests that the traditional Mott criterion $N_c \sim N_c^{(0)}$ should be replaced by a much larger value, but a definitive conclusion awaits more careful analysis, both theoretical and experimental.
Appendix \ref{sec:NclargeN} gives a complimentary estimate of $N_c$ from the conducting side, following a separate line of reasoning also suggested by Mott \cite{Mott1968}, and finds that $N_c$ is similarly greatly enhanced over Eq.\ (\ref{eq:Mottcrit}).

To estimate $N_c$ from the insulating side, consider that when a single electron is bound to an isolated donor impurity, it forms a neutral complex (a ``D$^0$ state'') with ionization energy $E_i$.  A second electron may also bind to the neutral D$^0$ complex, occupying a ``D$^-$ state'' with a significantly lower ionization energy.  For example, in conventional semiconductors such as GaAs the ionization energy of the D$^{-}$ state is only $\approx 0.1 E_i$. \cite{shklovskii_electronic_1984} One can therefore say that the difference in energy $U$ between the energy levels of the first electron and the second electron is of order $E_i$.

When two donor impurities are separated by a finite distance $R$, there is a nonzero amplitude for hopping between the D$^0$ state of one donor and the D$^0$ state of the other.  In cases where this amplitude is weak compared to $E_i$, the hopping is described by the hopping integral
\be 
\gamma(R) = \int d^3r \, \psi^*(r) \frac{e^2}{\epsilon r} \psi(|\r - \R|).
\ee
Here, $\psi(|\r - \R|)$ describes a D$^0$ wave function centered at the position $\r = \R$, and $e^2/(\epsilon r)$ describes the Coulomb potential created by a donor impurity at the origin.  Inserting the expression for the wave function from Eq.\ (\ref{eq:psi0}), and evaluating the integral in the limit $k_0 a_B \gg 1$, gives
\be 
\gamma(R) \simeq \frac{2 e^2}{\epsilon a_B } \frac{\sin(k_0 R)}{k_0 R} \ln(k_0 a_B) \exp\left[ - \frac{R}{a_B} \right]. 
\label{eq:gamma}
\ee
For separations $R \gg k_0^{-1}$, the large factor $k_0 R$ in the denominator of this expression implies that the hopping amplitude is substantially weaker than for the conventional case, where $\gamma(R) \sim (e^2/\epsilon a_B) \exp[-R/a_B]$.  This smaller amplitude is a consequence of the fast oscillation of the wave function [see Fig.\ \ref{fig:psi}(b)], which implies that the quantum-mechanical overlap between two D$^0$ wave functions can be relatively small even when their separation $R$ is shorter than $a_B$.

Mott and Davis \cite{Mott1980} suggested that the system undergoes an IMT when the donor impurity concentration $N_D$ is such that $\gamma(R = N_D^{-1/3})$ is comparable in magnitude to $U \sim E_i$.  Equating the results from Eq.\ (\ref{eq:EbMH}) and (\ref{eq:gamma}) suggests that $\gamma$ and $U$ become comparable only when $R \lesssim k_0^{-1}$.  This implies a critical concentration 
\be 
N_c \sim k_0^3.
\label{eq:Nc}
\ee 
For MVIs, where $k_0$ is of order of the inverse lattice constant $a_0$, this expression implies a huge critical concentration, larger than the Mott value by a factor $\sim (a_B/a_0)^3$ (which in SmB$_6$ is of order $10^3$). Since the IMT apparently occurs when the distance between impurities is of the order of the lattice constant, the precise value of $N_c$ will depend on atom-scale details that are beyond the continuum model we are using here.

\section{impurity band and dc conductivity}
\label{sec:DC}

Given the apparently huge critical concentration $N_c$, for the remainder of this paper we assume that the dopant concentration $N_D \ll N_c$, which is the condition normally referred to as ``light doping''.  We also assume the impurity band to be lightly, or incompletely compensated, so that the concentration of acceptors $N_A$ satisfies $(N_D - N_A)/N_A \gtrsim 1$.  At any nonzero concentration $N_A$ of acceptors, the donor impurity band is not completely filled, and in the limit of zero temperature the chemical potential is pinned to the donor impurity band.  

\begin{figure}[htb!]
\centering
\includegraphics[width=1.0 \columnwidth]{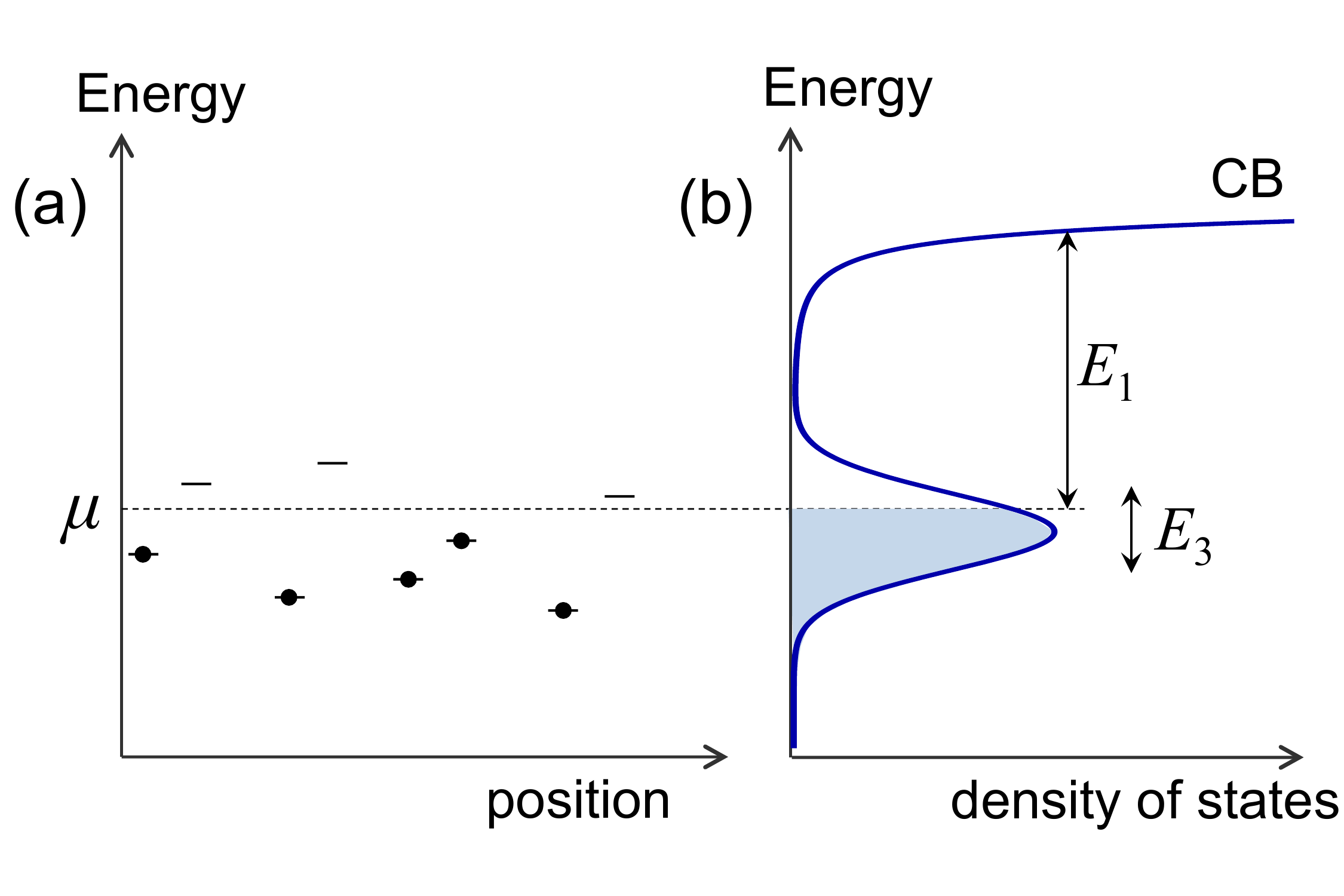}
\caption{The impurity band at light doping and incomplete compensation.
(a) Donor impurity energy levels, which have a random shift relative to the chemical potential $\mu$ due to Coulomb interactions with acceptors and empty donor sites.  (b) The density of states as a function of energy, showing the impurity band and the conduction band edge (CB). The energies $E_1$ and $E_3$ are defined by Eq.\ (\ref{eq:impuritybandsigmaT}). }
\label{fig:lightdoping}
\end{figure}

In this limit of light doping and incomplete compensation, one can think of the impurity band quasi-classically: as a partially-filled set of energy levels described by a probability distribution having mean energy $E_i$ below the bottom of the conduction band.
This distribution is illustrated schematically in Fig.\ \ref{fig:lightdoping}.  The on-site repulsion between electrons provides a large energy penalty for double occupation of a given donor ion, as discussed in the previous section, so that in the ground state essentially all impurity states (for monovalent donors) are either singly-occupied or empty.  Throughout this section and the next we focus only on the lowest energy level (the D$^0$ state for each impurity).

Were it not for the long-range Coulomb interactions between impurity states, all such impurity levels would have the same energy (neglecting the weak quantum overlap between impurity state wavefunctions).  However, the presence of negatively-charged, occupied acceptor sites and positively-charged, unoccupied donor ions creates a random Coulomb potential that shifts the energies of individual donor states.  This random potential gives the impurity band a finite width in energy, of order $e^2 N_D^{1/3}/\e$. \cite{shklovskii_electronic_1984} Thus, the density of (localized) states $g$ in the impurity band is of order
\be 
g \sim \frac{N_D}{e^2 N_D^{1/3}/\epsilon} = \frac{N_D^{2/3}\epsilon}{e^2}.
\label{eq:g}
\ee
The exact position of the chemical potential within the impurity band depends on the degree of compensation.  

In cases where the chemical potential is pinned to an impurity band, one can generically write the temperature-dependent conductivity at not-too-low temperature as 
\be 
\sigma(T) = \sigma_1 \exp\left[ - \frac{E_1}{k_B T} \right] + \sigma_3 \exp \left[ - \frac{E_3}{k_B T} \right],
\label{eq:impuritybandsigmaT}
\ee
where the prefactors $\sigma_1$ and $\sigma_3$ have only a power-law dependence on temperature.
The first term on the right-hand side of Eq.\ (\ref{eq:impuritybandsigmaT}) denotes the conductivity associated with activation of electrons from the impurity band to the conduction band, so that $E_1 \simeq E_i$.  The second term is associated with hopping conductivity among impurity band states, so that $E_3$ is determined by the impurity band width ($E_3 = 0.99 e^2 N_D^{1/3}/\e$). \cite{shklovskii_electronic_1984} Finite compensation (a finite concentration $N_A$ of acceptors) provides a small correction to $E_3$, of order $E_3 (N_A/N_D)^{1/4}$.

At sufficiently low temperature, $\kb T \ll E_3 (N_D a_B^3)^{1/3}$, Eq.\ (\ref{eq:impuritybandsigmaT}) gives way to variable-range hopping conduction, where $\sigma(T) \propto \exp[-(T_0/T)^\eta]$, with $T_0$ a constant and $\eta$ an exponent smaller than unity.  \cite{shklovskii_electronic_1984}
While activated behavior has been observed over many orders of magnitude in SmB$_6$ \cite{eo2018robustness}, with an activation energy in the range $2$--$4$\,meV, variable range hopping has not yet been observed.  
So far it has not been possible to probe the bulk transport below $T \approx 2$\,K, either because of shunting of the bulk conduction by metallic surface states \cite{fisk1, fisk2, syers_tuning_2015} or because of a very large value of the bulk resistance. \cite{eo2018robustness}  However, if these limitations are circumvented (say, by studying larger samples and/or by using magnetic impurities to gap out the surface states), then variable-range hopping transport should appear at sufficiently low temperatures.%
\footnote{In general the magnitude of variable-range hopping conductivity $\sigma(T)$ at a given temperature $T$ depends on the hopping integral $\gamma$.  Since Eq.\ (\ref{eq:gamma}) is distinct from the conventional semiconductor case, variable range hopping is in principle different for MVIs.  However, since $\gamma(R)$ has the same exponential factor $\exp[-R/a_B]$, only the power-law prefactor of the dependence $\sigma(T)$ is affected, while the exponential part of the dependence remains the same as in the usual semiconductor case.}

\section{AC (optical) conductivity}
\label{sec:AC}

When the chemical potential resides amid localized states, the zero-frequency conductivity vanishes at zero temperature.  At finite frequency $\omega$, however, an applied electric field can induce an electron to transition from a filled to an empty localized state.  The conductivity $\sigma(\omega)$ is therefore finite at $\omega > 0$ even at zero temperature.  The theory of this AC conductivity was worked out by Mott \cite{mott_conduction_1970, mott_electronic_1971} and by  Shklovskii and Efros \cite{shklovskii_zero-phonon_1981} for conventional semiconductor impurity bands. The latter showed that, at frequencies  such that $\hbar \omega \ll e^2/\epsilon \abo$, the AC conductivity
\be 
\sigma(\omega) = A \frac{e^4}{\epsilon} g^2 [\abo]^4 \omega,
\label{eq:sigmaomegaES}
\ee 
where $A$ is a dimensionless prefactor that has only a logarithmic dependence on frequency.  The factor $[\abo]^4$ arises because in conventional semiconductors the tunneling rate between localized states decays exponentially with their separation $r$ as $\sim \exp[-2r/\abo]$.

In our problem, however, the overlap between localized wave functions decays substantially even at distances much shorter than $a_B$, and so Eq.\ (\ref{eq:sigmaomegaES}) should be modified.  In order to derive the proper expression, one can follow the derivations of Mott \cite{mott_conduction_1970, mott_electronic_1971} and Shklovskii and Efros \cite{shklovskii_zero-phonon_1981} as follows.

Consider a time-dependent electric field $\vec{\mathcal{E}} = \mathcal{E}_0 \hat{z} [\exp(i \omega t) + \exp(-i \omega t)]/2$ applied in the $\hat{z}$ direction.  According to Fermi's golden rule, such an electric field introduces a transition from some state $|i\rangle$ to another state $|j\rangle$ with rate 
\be 
\Gamma_{ij} = \frac{2 \pi}{\hbar} \left| \left \langle i \left| \frac{e \mathcal{E}_0 z}{2} \right| j \right \rangle \right|^2 \delta(E_j - E_i - \hbar \omega),
\nonumber
\ee 
where $E_i$ and $E_j$ are the energies of states $i$ and $j$, respectively.
The quantity $-e \langle i|z|j\rangle$ is the transition dipole moment, which determines the rate of transition between states $i$ and $j$.  Let us denote $z_{ij} \equiv \langle i|z|j\rangle$.  Suppose that the states $i$ and $j$ are localized, and that the vector $\r_{ij}$ connecting their centers has a magnitude $r_{ij}$ and forms an angle $\theta_{ij}$ with the field.  Then
\be 
z_{ij}(r_{ij}, \theta_{ij}) = \int  z \psi^*(r) \psi(|\r - \r_{ij}|) d^3 r .
\ee 
Inserting the wave function from Eq.\ (\ref{eq:psi0}) into this expression gives, to within an overall dimensionless prefactor of order unity, 
\be
z_{ij}(r_{ij}, \theta_{ij}) \sim \frac{\cos \theta_{ij}}{k_0^2 r_{ij}} \left( \cos (k_0 r_{ij}) - \frac{\sin(k_0 r_{ij})}{k_0 r_{ij}} \right)
\ee 
at separations $r_{ij} \ll a_B$.  At larger separations $r_{ij} \gg a_B$, the value of $z_{ij}$ decays exponentially, $z_{ij} \propto \exp(-r_{ij}/a_B)$.

Let us denote by $G(E_i, E_j, r)$ the probability density per unit volume of the system for finding two states with energies $E_i$ and $E_j$ and separation $r$.  Then the total power absorbed per unit volume of the system is
\be 
P = \hbar \omega \int dE_i \int dE_j \int d^3 r \Gamma_{ij} G(E_i, E_j, r) f(E_i) [1 - f(E_j)].
\ee 
Here, $f(E)$ is the Fermi function, and the factor $f(E_i) [1 - f(E_j)]$ in the integrand corresponds to the probability that state $i$ is filled and $j$ is empty.  In the remainder of this section we will consider the case of zero temperature.  The optical conductivity $\sigma(\omega)$ can be found by calculating the power $P$ and equating it to $\sigma(\omega) \mathcal{E}_0^2/2$.

Mott originally assumed that $G(E_i, E_j, r) = g^2$, in effect assuming that the energies at sites $i$ and $j$ are independent.  Under this assumption the two integrals over energy yield only a factor $\hbar \omega$, corresponding to the range of energies near the chemical potential that can absorb a photon. \cite{mott_electronic_1971}  However, Shklovskii and Efros pointed out that when the sites $i$ and $j$ are not too far apart, the energy of the empty state $j$ is determined in part by its repulsion to the electron in filled state $i$. Consequently, even a state deep below the chemical potential can potentially absorb a photon if it happens to have an empty state close by, such that $E_j - E_j - e^2/(\epsilon r_{ij}) = \hbar \omega$.  Using this logic and evaluating the integrals over energy gives \cite{shklovskii_zero-phonon_1981}
\be 
\frac{2 P}{\mathcal{E}_0^2} = \sigma(\omega) = \pi e^2 g^2 \omega \int d^3 r \left( \hbar \omega + \frac{e^2}{\epsilon r} \right) [z_{ij}(r, \theta)]^2.
\label{eq:sigmaomegaint}
\ee
This integral should be taken only over $r$ larger than a particular cutoff value $r_\omega$, defined by $2I(r_\omega) = \hbar \omega$, where $I(r)$ is the overlap integral between two states with separation $r$.  (The overlap integral $I(r)$ is of the same order as the hopping integral $\gamma(r)$. \cite{shklovskii_electronic_1984})  States with $r < r_\omega$ are strongly hybridized, and their energy splitting arising from hybridization is larger than $\hbar \omega$.  Inserting the expression for $z_{ij}$ and performing the integration over $r$ gives, up to an overall numerical coefficient,
\be 
\sigma(\omega) \sim \frac{e^4 g^2 \omega}{\epsilon k_0^4} \ln(a_B/r_\omega).
\label{eq:sigmaomega}
\ee 
This expression is valid at frequencies $\hbar \omega \ll e^2/(\epsilon a_B)$.

Thus we arrive at a striking result: the optical conductivity $\sigma(\omega)$ resembles the usual result for semiconductors, but with the decay length $\abo$ replaced by the much shorter length $k_0^{-1}$.  Intuitively, one can think that this replacement comes because an electron cannot absorb a photon unless it has significant overlap with another state, and in our problem only states with separation comparable to $k_0^{-1}$ have strong overlap.

For systems where $k_0$ is of the order of the inverse lattice constant, including the MVIs discussed so far, one can write Eq.\ (\ref{eq:sigmaomega}) very simply in terms of the doping fraction $x = N_D a_0^3$.  In particular, substituting the expression for the density of states of the impurity band [Eq.\ (\ref{eq:g})] gives the simple expression
\be 
\sigma(\omega) \sim \epsilon x^{4/3} \omega.
\ee 
Assuming $\epsilon$ on the order of $10^3$ and the doping level $x$ on the order of a few percent \cite{phelan_chemistry_2016} gives $\sigma(\omega) \sim 10$\,$\Omega^{-1} \textrm{cm}^{-1} \textrm{THz}^{-1}$.  This is consistent with what is seen in experiments on SmB$_6$ for frequencies $\omega \lesssim 1$\,THz. \cite{laurita_anomalous_2016}  At larger frequencies $\sigma(\omega)$ is presumably dominated by the excitation of localized electrons to delocalized states in the conduction band.

\section{Specific heat}
\label{sec:CV}

Unlike the transport properties considered in the previous sections, the specific heat $C_V$ is a thermodynamic property, and in the limit of light doping it is therefore largely insensitive to the quantum mechanical overlap between wave functions.  One can therefore largely recapitulate results for the classical semiconductor impurity band.  The primary difference for the case of MVIs is that the system can tolerate a much larger impurity concentration within the impurity band, as explained in Sec.\ \ref{sec:IMT}.

Generally speaking, in a lightly-doped semiconductor at low temperature, the specific heat arising from donor electrons has two contributions, corresponding to excitations in the spin and the charge sectors.  Here we will mostly consider the latter class of excitations, which is associated with thermal excitation of electrons from filled states to empty states within the impurity band.  More precisely, the low energy charge excitations correspond to simultaneous rearrangement of multiple electrons within the impurity band.  At small energy these excitations are classical in nature and correspond to a simultaneous rearrangement of many electron-hole pairs.  The properties of these excitations were studied by Baranovskii, Shklovskii, and Efros \cite{baranovskii_elementary_1980}, who found that the density of states for excitations with energy $E$ is $\Phi(E) \sim g \ln^{1/2} [e^6 g/(\e^3 E)]$, where $g$ is the density of states in the impurity band [see Eq.\ (\ref{eq:g})].  The logarithmic depletion of the density of states $\Phi$ for many-particle rearrangement at small $E$ arises from  the mutual interaction between compact ``dipolar" electron-hole excitations.

Using this expression for the density of states, one can calculate the specific heat as
\be 
C_V = \frac{d}{d T} \left[ \int E \Phi(E)  \exp\left( -\frac{E}{\kb T} \right) dE \right].
\label{eq:CVintegral}
\ee
The quantity in brackets in Eq.\ (\ref{eq:CVintegral}) represents the total electronic energy per unit volume relative to the ground state.  Evaluating this expression gives \cite{baranovskii_thermodynamic_1982}
\be 
C_V \sim \frac{\kb^2 \e N_D^{2/3}}{e^2} T \ln^{1/2} \left( \frac{e^2 N_D^{1/3}}{\e \kb T} \right).
\label{eq:CVcharge}
\ee
For SmB$_6$, where $\epsilon$ is of order $10^3$ and the lattice constant $a_0 = 4.13$\,$\ang$, this expression gives $C_V/T \sim 100 x^{2/3}$\,mJ/(mol K$^2$), where $x = N_D a_0^3$ is the doping fraction, which is presumably on the order of a few hundredths.  
For comparison, experimental studies of SmB$_6$ report a specific heat ranging between $4$ and $10$\,mJ/(mol K$^2$), with a value that increases upon intentional doping. \cite{SS15, fuhrman_screened_2017}

Theoretically, the linear-in-$T$ specific heat in the charge sector of a doped semiconductor can be expected to persist with increasing temperature until $\kb T$ becomes as large as the width of the impurity band.  At larger $T$ the specific heat is dominated by activation of electrons from the impurity band to the conduction band, which has an exponential dependence on temperature with an activation energy $E_i$. 

Let us now briefly comment on the contribution to the specific heat associated with excitations in the spin sector. In the simplest description, each impurity state is considered to be spin-degenerate and noninteracting.  In such a situation there is no spin contribution to the specific heat.  However, a given pair of donor electrons with finite separation have an antiferromagnetic interaction with each other, and this interaction contributes to the specific heat when the concentration of donor impurities is not too low. \cite{Bhatt_magnetic_1986} In particular, Bhatt and Lee showed that for conventional semiconductors at sufficiently low temperature the antiferromagnetic interaction leads to ``frozen" clusters of hybridized, random-singlet-like spins. \cite{LeeBhatt1} The size of these clusters grows with decreasing temperature, leading to an unusual power-law dependence of the specific heat on temperature
\be 
C_V \propto T^{1-\alpha},
\label{eq:CVspin}
\ee 
with $0 < \alpha < 1$.  \cite{LeeBhatt2}
In principle, this contribution to the specific heat from the spin sector dominates over the charge sector contribution when the temperature is low enough.  However, the strength of the exchange interaction $J$ between two impurity electrons declines strongly with their separation $r$.  In conventional semiconductors $J \propto \exp[-2r/a_B]$, so that Eq.\ (\ref{eq:CVspin}) is realized only below some temperature scale that is exponentially small in the parameter $1/(N_D^{1/3} a_B)$.  Thus, at light doping and realistic temperatures the spin contribution can likely be neglected relative to the charge sector contribution [Eq.\ (\ref{eq:CVcharge})].
The generalization of this random singlet physics to the case of MVI impurities, where the exchange interaction has only a power-law decay at distances $r \ll a_B$, remains to be explored.

\section{Summary and conclusion}
\label{sec:conclusion}

In this paper we have considered the properties of donor and acceptor impurities in MVIs from the perspective of doped semiconductors.  While the properties of semiconductor impurity bands have been well studied for many decades, the unusual band structure arising from hybridization between the light $d$- and heavy $f$-bands in MV insulators has deserved special attention, and has led to the modification of a number of results. 

In particular, we find that charged impurities in SmB$_6$ or other MVIs can naturally lead to mid-gap impurity states with an unusual ``one-dimensional hydrogen atom"-like ionization energy.  For SmB$_6$ this solution implies an ionization energy in the range $1-5$\,meV and an effective Bohr radius $\sim 60$\,$\ang$.  Despite this relatively large Bohr radius and low ionization energy, our estimates suggest that doping does not produce an insulator-to-metal transition at doping concentration $N_D \sim 1/a_B^3$.  Instead, the rapidly-oscillating nature of the impurity wave function leads to a low degree of quantum-mechanical overlap between impurity states, so that the insulator-to-metal transition happens only at a much higher level of doping $N_c \sim k_0^3$, which is potentially as large as an order-unity constant times $1/a_0^3$.  We have also shown that the impurity band exhibits a linear-in-$\omega$ optical conductivity, with a coefficient that is strongly different from the conventional semiconductor case, and a linear-in-$T$ specific heat.  Our results are summarized in Table \ref{tab:impurityband}.

\begin{table}[tb!]
\begin{center}
    \begin{tabular}{ | c | c |}
    \hline
    \thead{Property} &   \thead{Behavior}  \\ \hline\hline
  \makecell{ionization energy} & $E_i \sim E_i^{(0)} \ln^2[k_0 \abo]$, \ Eq.\ (\ref{eq:EbMH})   \\   \hline
  \makecell{critical doping for IMT} & $N_c \sim k_0^{3}$ , \  Eq.\ (\ref{eq:Nc}) \\   \hline
\makecell{DC  conductivity} & \ $\sigma \propto \exp[-E_i/\kb T]$, \ Eq.\ (\ref{eq:impuritybandsigmaT}) \   \\ \hline
   \makecell{AC conductivity} & $\sigma \propto \omega$, \ Eq.\ (\ref{eq:sigmaomega})   \\ \hline
   \makecell{specific heat} & $C_V \propto T$, \ Eq.\ (\ref{eq:CVcharge})  \\ \hline
    \end{tabular}
    \caption{ Summary of properties arising from the dopant impurity band.}
    \label{tab:impurityband}
\end{center}
\end{table}

While the results we have presented seem, so far, consistent with experimental results, it is worth emphasizing that we have not attempted to explain the most dramatic experimental feature, which is the appearance of bulk quantum oscillations in the magnetization.  Put bluntly, we see no mechanism by which the impurity band alone can produce such oscillations.  
It may be that impurity-independent theoretical proposals such as those of Refs.\ \onlinecite{DC, Baskaran, ColemanSC, CooperMB, CooperExc, FWMB} are necessary to explain quantum oscillations, while other ``gapless'' features are explainable in terms of an impurity band.  It is worth noting, though, that the mechanism suggested by Ref.\ \onlinecite{CooperMB} is enhanced when the chemical potential is pinned closer to the band edge (as at an impurity band). The suggested explanation of Ref.\ \onlinecite{FuQO} also relies crucially on the existence of in-gap impurity states, which are assumed in Ref.\ \onlinecite{FuQO} but not derived in detail.  (A recent experimental study has also suggested that bulk quantum oscillations in flux-grown SmB$_6$ samples may arise from
embedded aluminum inclusions. \cite{thomas2019quantum})

Further experiments can help to confirm or refute the results we have presented here, for example, by further studying the bulk transport at low temperature, or by detailed studies of conductivity and specific heat as a function of doping.

\acknowledgments

My thanks for this work go primarily to D.\ Chowdhury, who played the role of my advisor on this project.
I am also grateful to P.\ Armitage, W.\ Fuhrman, Y.\ S.\ Eo, C.\ Kurdak, K.\ Sun, A.\ Rosch, and B.\ I.\ Shklovskii for useful discussions. 
The work was supported by the NSF STC ``Center for Integrated Quantum Materials" under Cooperative Agreement No.\ DMR-1231319. 
Part of this work was performed at the Aspen Center for Physics, which is supported by National Science Foundation grant PHY-1607611.

\bibliography{SmB6}

\widetext 

\appendix

\section{Calculation of the dielectric constant from the hybridized two-band model}
\label{sec:epsilon}

Here we show how to calculate the electronic part of the dielectric constant  $\epsilon$ using our model Hamiltonian for MV insulators, and we show that our result is consistent with known values for SmB$_6$.

The general expression for the dielectric function is \cite{mahan_many-particle_1990}
\be 
\e(\q) = 1 + \frac{4 \pi e^2}{q^2} \sum_{\k} 2 \left\lvert \langle \k, v | \exp(i \q \cdot \r ) | \k + \q, c \rangle \right\rvert^2 \frac{f(\k, v) - f(\k + \q, c)}{E_{+}(\k + \q) - E_{-}(\k)}.
\label{eq:epsqintegral}
\ee
Here, $\q$ is the wave vector, $|\k, c\rangle$ and $|\k, v\rangle$ represent the momentum eigenstates of the conduction and valence bands, respectively, and $E_{\pm}(\k)$ are the conduction and valence band dispersion relations, given by Eq.\ (\ref{eq:Epm}).  $f(\k, c)$ and $f(\k, v)$ are the Fermi functions describing the conduction and valence bands; at zero temperature, $f(\k, v) - f(\k + \q, c) = 1$.  In the limit of small $q$, one can replace the sum over $\k$ with an integral, $\sum_\k \rightarrow \frac{a_0^3}{(2 \pi)^3} \int d^3 \k$.

The conduction and valence band eigenstates are found by diagonalizing the Hamiltonian 
\be 
H = \frac{E_d + E_f}{2} + 
\begin{bmatrix}
\frac12(E_d - E_f) & V \\
V & -\frac12(E_d - E_f)
\end{bmatrix},
\ee 
which gives the eigenvalues from Eq.\ (\ref{eq:Epm}) and the eigenstates
\be 
|\k, c \rangle = 
\begin{bmatrix}
u_{+} \\
v_{+}
\end{bmatrix} ,
|\k, v \rangle = 
\begin{bmatrix}
u_{-} \\
v_{-}
\end{bmatrix} ,
\ee
with 
\begin{eqnarray}
u_\pm & = & \sqrt{\frac{E_+ \pm (E_d - E_f)/2}{2 E_{+}}}, \\
v_\pm & = & \pm \sqrt{\frac{E_{+} \mp (E_d - E_f)/2}{2 E_{+}}}.
\end{eqnarray}
Here we use for the $d$ and $f$ band dispersions
\begin{eqnarray}
E_d(\k) & = & -2 t_d \left[ \cos(k_x a_0) + \cos(k_y a_0) + \cos(k_z a_0) \right] \nonumber \\
E_f(\k) & = & 2 t_f \left[ \cos(k_x a_0) + \cos(k_y a_0) + \cos(k_z a_0) \right].
\end{eqnarray}
The nearest neighbor hopping elements $t_d$ and $t_f$ are related to the band masses by $t_{d,f}\approx\hbar^2/(2 m_{d,f} a_0^2)$.

The coherence factor can now be calculated by evaluating the inner product in Eq.\  (\ref{eq:epsqintegral}) and performing the integral over $\k$ numerically.  If one takes for the band masses $m_d = 1.5 m_0$ and $m_f = 50 m_0$, then the corresponding values of the hopping matrix elements are $t_d \approx 150$\,meV and $t_f \approx 4.5$\,meV, and the choice $V = 15$\,meV gives $\e \approx 1600$ in the limit of $q \rightarrow 0$.

\section{Variational estimate of the impurity ionization energy}
\label{sec:variation}

An alternate way to estimate the impurity state energy is to use a variational approach, which gives an upper bound for the energy of the state, and therefore a lower bound for the ionization energy. Here we use as an ansatz the cigar-shaped, hydrogen-like wave function
\be 
\psi(\r) = \frac{\exp\left[-\sqrt{\eta^2 r_\parallel^2 + r_\perp^2}/b_\perp \right] }{ \sqrt{\pi b_\perp^3/\eta^2} } \exp[i \k_0 \cdot \r ],
\ee
where $\k_0$ is an arbitrarily-chosen point in momentum space along the conduction band minimum [see Fig.\ \ref{fig:dispersion}(b)].  We write the wave function in cylindrical coordinates, so that $r_\perp$ is the distance in real space from the impurity center along the direction perpendicular to the minimum surface, and $r_\parallel$ is the distance along the parallel direction.  For simplicity, we take a point $\k_0 \parallel  (\hat{x} + \hat{y} + \hat{z})$.  The variables $b_\perp$ and $\eta$ are variational parameters, such that $b_\perp$ represents the wave function decay length along the perpendicular direction and $\eta > 1$ is the wave function anisotropy.

The Fourier transform of the variational wave function is given by
\be 
\widetilde{\psi}(\k) = \sqrt{ \frac{64 \pi b_\perp^3}{\eta^2}} \frac{1}{[1 + b_\perp^2(k_\perp^2 + k_\parallel^2/\eta^2)]^2},
\ee
with $k_\perp = (\k - \k_0)\cdot \hat{n}$ being the wave vector component relative to $\k_0$ along the surface normal direction $\hat{n} = (\hat{x} + \hat{y} + \hat{z})/\sqrt{3}$, and $k_\parallel^2 = |\k - \k_0|^2 - k_\perp^2$.

The Coulomb energy of the variational state can be written
\be
E_c(b_\perp, \eta) = - \int d^3 \r \left\lvert \psi(\r) \right\rvert^2 \frac{e^2}{\epsilon |\r|}.
\label{eq:Ecvar}
\ee
In the limit where the ionization energy $E_i$ is much smaller than the band gap $E_g$, only the conduction band is relevant for the kinetic energy $E_k$ of the variational state.  So one can write
\be 
E_k(b_\perp, \eta) = \int \frac{d^3 \k}{(2 \pi)^3} E_{+}(\k) \left\lvert \widetilde{\psi}(\k) \right\rvert^2.
\label{eq:Ekvar}
\ee 
These two integrals can be evaluated numerically for a generic choice of $b_\perp$ and $\eta$.

The variational estimate for the ionization energy is 
\be 
E_i \approx E_{+}(\k_0) - \underset{b_\perp, \eta}{\textrm{min}} \left[ E_c(b_\perp, \eta) + E_k(b_\perp, \eta) \right].
\ee 
The first term on the right-hand side of this expression subtracts defines the impurity state energy relative to the bottom of the conduction band, $E_{+}(\k_0) = E_g/2$.

The integrals from Eqs.\ (\ref{eq:Ecvar}) and (\ref{eq:Ekvar}) can be evaluated numerically, and the resulting sum can be minimized numerically over both variational parameters.  For SmB$_6$, using $V = 15$\,meV, $m_d = 1.5 m_0$, $m_f = 50m_0$ and $\epsilon = 600$, this procedure gives  $E_i \approx 0.8$\,meV.  The corresponding values of the variational parameters are $b_\perp \approx 49~\ang$ and $\eta = 6.9$.

\section{Estimate of the critical concentration for the IMT from the conducting side}
\label{sec:NclargeN}

\subsection{Existence of an impurity bound state}

Mott and others \cite{Mott1968} suggested that the IMT can be thought about from the metallic side as follows.  When the doping is heavy enough that $N_D \gg N_c$, one can consider that the Fermi level is well above the bottom of the conduction band and the electron concentration is mostly uniform spatially.  In this case the Coulomb potential created by individual donors is screened by itinerant electrons over a distance given by the Thomas-Fermi screening radius $r_s$.  If this screening radius is sufficiently short, then the Coulomb potential of a single donor does not admit a bound electron state.  However, as the electron concentration is reduced, the screening radius $r_s$ grows, and at a critical doping $N_c$ it becomes possible to make a bound state of an electron to a single donor. One can take this value of $N_c$ as an estimate for the concentration at the IMT.

Let us now assume the existence of such a metallic state in an MVI and ask under which conditions a single donor impurity can create a strong enough potential to localize an electron.

When there is no screening, the Coulomb potential $V_C(r)$ created by a charged impurity is $V_C(r) = e^2/(\epsilon r)$.  In situations with a sufficiently large concentration of itinerant electrons, however, the electric potential is truncated by Thomas-Fermi screening, and takes the form
\be 
V_C(r) = \frac{e^2}{\epsilon r} \exp[-r/r_s].
\label{eq:Yukawa}
\ee
The value of $r_s$ depends in general on the electron density; as we discuss in Appendix \ref{sec:screeninglowN}, its minimal value for our problem is $r_s \sim \left(k_0^{2} \abo \right)^{-1/3} \gg k_0^{-1}$.  Let us for the moment keep $r_s$ as a variable, and we will determine how small $r_s$ must be in order to preclude the existence of an electron bound state.

A direct solution of the Schr\"{o}dinger equation with the potential (\ref{eq:Yukawa}) is a difficult problem, but we can take a variational approach by examining the expectation value of the energy of the trial wave function
\be 
\psi_\text{tr}(r; b) = \frac{1}{\sqrt{\pi b}} \frac{\sin(k_0 r)}{r} \exp\left[- \frac{r}{b} \right].
\ee
This wave function closely resembles the solution to the Schrodinger equation for the unscreened Coulomb potential [Eq.\ (\ref{eq:psi0})], except that the exponential decay length $b$ has been left as a variational parameter.  The energy $E_\text{tr}(b)$ of this trial wave function represents an upper bound for the ground state energy of an electron interacting with the screened potential $V(r)$, so that if $E_\text{tr}(b) < 0$ for any finite value of $b$ then there exists a bound state.

Let us separate the energy $E_\text{tr}(b)$ into kinetic and potential energy parts, such that $E_\text{tr}(b) = K_\text{tr}(b) + U_\text{tr}(b)$.  The kinetic energy part, relative to the conduction band bottom, is
\begin{align}
K_\text{tr}(b) & = \int \frac{d^3 k}{(2 \pi)^3} E_{+}(\k) \left| \widetilde{\psi}_\text{tr}(\k) \right|^2 \nonumber \\
& \simeq \frac{\hbar^2}{2 m b^2},
\label{eq:Ktr}
\end{align}
where $\widetilde{\psi}_\text{tr}(\k)$ is the Fourier Transform of $\psi_\text{tr}(r)$, and the second equality is taken in the limit $b k_0 \gg 1$.  The potential energy contribution to $E_\text{tr}$ can be written
\be 
U_\text{tr}(b) = \int d^3 r \, V_C(r) \left|\psi_\text{tr}(r; b) \right|^2.
\ee
The dominant contribution to this integral comes from distances $r$ such that $k_0^{-1} \ll r \ll r_s$, over which the Coulomb potential is essentially unscreened, $V(r) \propto 1/r$, and the electron density $|\psi_\text{tr}|^2 \propto 1/r^2$.  One therefore gets
\be 
U_\text{tr}(b) \simeq - \frac{2 e^2}{\epsilon b} \ln (k_0 r_s).
\label{eq:Utr}
\ee
Minimizing the total energy $E_\text{tr} = K_\text{tr} + U_\text{tr}$ with respect to $b$ gives $b = \abo/[2 \ln (k_0 r_s)]$, and
\be 
\min_{b > 0} \, E_\text{tr}(b) \simeq -\frac{2 e^2}{\epsilon \abo} \ln^2 ( k_0 r_s ).
\ee

Thus we arrive at the conclusion that $E_\text{tr}$ is negative, and therefore bound states exist, any time the screening radius is longer than $k_0^{-1}$.  Such short screening radii that $k_0 r_s < 1$ are not possible within the assumptions of our model, since $k_0^{-1}$ is of the same order as the lattice constant.  The analysis therefore suggests that the system is in the insulating state at all doping concentrations $N_D \lesssim k_0^3$, at which our description of the conduction band is valid.

\subsection{Screening of the Coulomb potential}
\label{sec:screeninglowN}

Here we discuss the screening of the Coulomb potential in a metallic system with a dispersion relation given by Eq.\ (\ref{eq:MHdispersion}), which is our model for the low-lying conduction band states of a MV insulator.  
Due to the large degeneracy of the conduction band minimum, the density of states in the conduction band diverges near the band edge as 
\be 
\nu(E) = \frac{\sqrt{2}}{\pi^2} k_0^2 \sqrt{\frac{m}{\hbar^2 E}},  
\label{eq:nuMH}
\ee
where $E$ is the energy relative to the band edge.  This divergence of the density of states at low energy is a crucial difference as compared to conventional semiconductors,\footnote{Equation (\ref{eq:nuMH}) suggests that the problem of critical doping in MV insulators is closely analogous to the problem of critical doping in a semiconductor in the extreme quantum limit of magnetic field. \cite{shklovskii_localization_1973} In this latter problem the density of states has a similar $\sim 1/\sqrt{E}$ divergence, and the resulting critical doping value $N_c$ is similarly much larger than the zero-field value.}
for which the density of states vanishes near the band edge as $\sim \sqrt{E}$.
The corresponding Fermi energy
\be 
E_F = \frac{\pi^4}{8} \frac{\hbar^2 N^2}{m k_0^4},
\ee
where $N$ is the concentration of conduction band electrons.
The Fermi surface takes the shape of a thin spherical shell, with radius $k_0$ and thickness $2k_F = \pi^2 N/k_0^2$.  So long as $N \ll k_0^3 \sim 1/a_0^3$, the thickness $k_F$ of the shell is much less than the radius $k_0$.

Screening of the Coulomb potential in a good metal is usually described by the Thomas-Fermi (TF) approximation, which gives the Yukawa potential of Eq.\ (\ref{eq:Yukawa}).  Its Fourier transform $\widetilde{V}_C(q)$ is
\be 
\widetilde{V}_C(q) = \frac{4 \pi e^2}{q^2 + r_s^{-2}},
\label{eq:VTF}
\ee
where $q$ is the modulus of the wave vector and 
\be 
r_s = \sqrt{\frac{4 \pi \epsilon}{e^2 \nu}} = \pi^{5/2} \sqrt{\frac{N \abo}{k_0^4} }. 
\label{eq:rs}
\ee
is the TF screening length.  Notice that as the concentration $N$ of itinerant electrons is reduced the screening length $r_s$ becomes shorter, owing to the rising density of states at low energy.  On the other hand, the Fermi wavelength $\lambda_F = 2 \pi/k_F = 4 k_0^2/(\pi N)$ becomes longer at low density.  The TF approximation is valid only when $r_s/\lambda_F \gg 1$, which implies that the TF description fails at densities $N \ll \left( k_0 \abo \right)^{8/3}/\left(\abo \right)^3$.  

At lower densities one can describe screening of the Coulomb potential using the static, momentum-dependent polarization function
\be 
\Pi(\q) = \int \frac{d^3 \k}{(2 \pi)^3} \frac{f(E(\k)) - f(E(\k + \q))}{E(\k) - E(\k + \q)},
\label{eq:polarization}
\ee
where $f(E)$ is the Fermi distribution function, which we consider at zero temperature.  The screened Coulomb potential is given by
\be 
\widetilde{V}_C(q) = \frac{4 \pi e^2}{\epsilon q^2 \left( 1 + \frac{4 \pi e^2}{\epsilon q^2} \Pi(q) \right)}.
\label{eq:Vqscreened}
\ee
For low enough momenta that $q  \ll k_F$, Eq.\ (\ref{eq:polarization}) gives $\Pi \simeq -\nu$, which implies that the screened potential is well-described by the usual TF result [Eq.\ (\ref{eq:VTF})] at wave vectors much smaller than $k_F$, or in other words at distances much longer than $\lambda_F$.  At $q \gg k_F$, on the other hand, evaluating the integral in Eq.\ (\ref{eq:polarization}) gives
\be 
\Pi(q) \sim -k_0^2 k_F  \times \frac{1}{\hbar^2 q^2/m}.
\label{eq:PinonTF}
\ee 
The dimensionless quantity $(4 \pi e^2/\epsilon q^2)\Pi(q)$ in the denominator of Eq.\ (\ref{eq:Vqscreened}) is therefore $\sim (k_F^2/r_s^2)/q^4$.  This quantity becomes large compared to unity when $q \ll \sqrt{k_F/r_s}$, which implies that there is significant screening of the potential at distances longer than some length scale $r_0$ defined by
\be 
r_0 \sim \sqrt{\frac{r_s}{k_F}} \sim \left( \frac{\abo}{N} \right)^{1/4}.
\label{eq:r0}
\ee
The quantity $r_0$ can therefore be thought of as an effective screening length at low enough electron concentrations that the TF approximation is no longer applicable. At such low concentrations $r_0$ is shorter than the Fermi wavelength, but is still much longer than $k_0^{-1}$.  Reducing the electron concentration $N$ causes $r_0$ to grow longer.

The minimal value of the screening length therefore occurs when $N \sim \left( k_0 \abo \right)^{8/3}/\left(\abo \right)^3$, at which point the lengths $r_0$ and $r_s$ coincide, and one obtains
\be 
r_s^{(\textrm{min})} \sim \frac{1}{\left(k_0^2 \abo \right)^{1/3}} \gg k_0^{-1}.
\ee

\end{document}